\documentclass[sn-mathphys,Numbered]{sn-jnl}

\usepackage{graphicx}%
\usepackage{multirow}%
\usepackage{amsmath,amssymb,amsfonts}%
\usepackage{amsthm}%
\usepackage{mathrsfs}%
\usepackage[title]{appendix}%
\usepackage{xcolor}%
\usepackage{textcomp}%
\usepackage{manyfoot}%
\usepackage{booktabs}%
\usepackage{algorithm}%
\usepackage{algorithmicx}%
\usepackage{algpseudocode}%
\usepackage{listings}%
\usepackage{longtable}
\usepackage{tabularx}
\usepackage{placeins}
\usepackage{natbib}

\begin{document}

\title[Article Title]{The Utility of Large Language Models and Generative AI for Education Research}


\author*[1]{\fnm{Andrew} \sur{Katz}}\email{akatz4@vt.edu}

\author[1]{\fnm{Umair} \sur{Shakir}}\email{irs933@vt.edu}

\author[1]{\fnm{Benjamin} \sur{Chambers}}\email{bdc0112@vt.edu}

\affil*[1]{\orgdiv{Department of Engineering Education}, \orgname{Virginia Tech}, \orgaddress{\city{Blacksburg}, \postcode{24060}, \state{VA}, \country{USA}}}




\abstract{\textbf{Background:} The use of natural language processing (NLP)
techniques in engineering education can provide valuable insights into
the underlying processes involved in generating text. While accessing
these insights can be labor-intensive if done manually, recent advances
in NLP and large language models have made it a realistic option for
individuals.
 
\textbf{Purpose:} The purpose of this paper is to present a set of
methods for utilizing NLP and large language models to thematically
analyze unstructured text data in engineering education.
 
\textbf{Design/Method:} This study explores and evaluates a combination
of clustering, summarization, and prompting techniques to analyze over
1,000 student essays in which students discussed their career interests.
The specific assignment prompted students to define and explain their
career goals as engineers. Using text embedding representations of
student responses, we clustered the responses together to identify
thematically similar statements from students. The clustered responses
were then summarized to quickly identify career interest themes. We also
used a set of a priori codes about career satisfaction and sectors to
demonstrate an alternative approach to using these generative text
models to analyze student writing.

\textbf{Results:} The results of this study demonstrate the feasibility
and usefulness of NLP techniques in engineering education research. By
automating the initial analysis of student essays, researchers and
educators can more efficiently and accurately identify key themes and
patterns in student writing. This can help to better understand student
learning and career interests, as well as inform teaching strategies and
program development.

\textbf{Conclusions:} This paper contributes to the growing body of
research on NLP in engineering education and provides a practical guide
for those interested in applying these techniques to their own research
or teaching contexts. The methods presented in this paper have broader
applications for engineering education and research purposes beyond
analyzing student essays. By explaining these methods to the engineering
education community, readers can utilize them in their own contexts.}

\keywords{engineering education, generative AI, qualitative analysis, student writing}



\maketitle

\section{Introduction}\label{sec1}

In engineering education research, qualitative data are ubiquitous, from student assignments and reflections to teacher notes and audio transcripts of interviews or group work. While qualitative data provide abundant information about teaching and learning, analyzing these data can be challenging and time-consuming without additional resources. For example, an instructor may teach four sections of 70 students each and assign a reflective writing exercise. Alternatively, a researcher may
collect open-ended survey responses from several thousand participants. How do they analyze those data at scale? Often, valuable insights from those data go unobserved due to a lack of efficient analysis methods. However, recent advances in natural language processing (NLP) and large
language models (LLMs) offer promising solutions to these challenges by automating text analysis and improving the accuracy and efficiency of theme identification \cite{devlin_bert_2019, radford_learning_2021}.

As a subfield of artificial intelligence, NLP techniques allow computers to understand, interpret, and generate human language. Powerful foundation language models that provide the base upon which other models can be fine-tuned \cite{bommasani_opportunities_2022,zhou_comprehensive_2023}, such as GPT-3 and BERT, can generate responses as if they
understand complex language patterns and generate relevant responses \cite{brown_language_2020, devlin_bert_2019}. These technologies have been applied successfully in various educational contexts, including analyzing student writing, assessing learning, and identifying themes in
qualitative data \cite{liu_automated_2022,camus_investigating_2020,ahmad_automatic_2022}. While NLP is not new, recent advances like the increasingly larger GPT models \cite{devlin_bert_2019, radford_language_2019} offer feasible ways for educators and
researchers to leverage these models' capabilities to analyze qualitative data at scale. The outline for this paper is as follows: we start by reviewing some of the background of NLP, LLMs, and applications
of NLP in education settings. We then explain our approaches to using LLMs and generative text models to analyze student writing in the Methods section. In the Results section, we present representative output from this process along with evaluations before ending with a discussion of what these results mean and limitations of these methods. 

\section{Relevant Background}\label{rel-background}
\subsection{A Brief Overview of Natural Language Processing}

Natural language processing is a subfield of artificial intelligence (AI) and linguistics that aims to enable computers to understand, analyze, interpret, and generate human language \cite{jurafsky_vector_2019}. Common NLP techniques process natural language data, such as text
and speech, for applications including machine translation, sentiment analysis, and information extraction \cite{manning_foundations_1999}. The field's origins can be traced back to the 1950s, with early work on machine translation and formal grammar \cite{mccarthy_proposal_2006, hutchins_example-based_2005}. Since then, approaches have evolved from rule-based (manually crafted rules and expert knowledge) to statistical
(machine learning and large datasets) and recently to deep learning (neural networks) \cite{goodfellow_deep_2016}.

During the 1950s-1980s, NLP tools were rule-based and deterministic in that they primarily used on grammar or vocabulary rules. These rule-based NLP approaches relied on a relatively small set of predefined rules and a limited vocabulary. Researchers encoded rules of human
language into algorithms, a process that could be time-consuming and often brittle given the broad diversity of language. From 1980-2000, a boom in computational capacities gave rise to statistics-based NLP approaches. These approaches relied on hand-crafted linguistic features
to quantify patterns and relationships in text corpora and make statistical inferences for performing NLP tasks. Examples of these NLP tools include Hidden Markov Models and Decision Trees, which had been used for parts-of-speech tagging, text parsing, and named entity recognition \cite{pulman_automatic_2005,graesser_coh-metrix_2004,attali_automated_2004}. The heavy reliance of these statistical NLP tools on hand-crafted rules limited their ability to handle anything outside of specific text structures and content. This was a non-trivial limitation.

In the period from 2000-2010, with more leaps in computational performance allowing for the analysis of readily available large digital text corpora, NLP tools began to incorporate machine learning (ML) concepts to train models with human-annotated data, a process known as supervised learning. Examples of these techniques included Regression, Support Vector Machines, Random Forests, Naïve Bayes classifiers, and n-gram representations. These ML-based NLP approaches mainly analyzed text or were able to quantify important information about the meaning of a sentence based on regular expressions or parse trees, and they used syntactic features (that describe the roles and dependencies of words in a sentence). However, these approaches were still inflexible in capturing the semantic meanings of words (relationships between words in a sentence).

That brings us to the modern period of deep learning--using neural networks with multiple layers. During this period, loosely marked from 2010 to the present, a major milestone in the field of NLP has been the development of word embedding methods, such as Word2Vec and GloVe \cite{mikolov_efficient_2013,pennington_glove_2014}. These embedding models consist of techniques for mapping words or phrases from a vocabulary into continuous vector
representations, capturing semantic meaning and context. Once in this vector space representation, one can use mathematical operations for analyzing text such as cosine similarity or clustering for identifying similar writing. The vector representations are also common inputs for
modern neural network-based NLP models.

Prior to the current generation of model architectures, recurrent neural networks (RNNs) were a popular form of deep learning model used in NLP to process sequential data, such as sentences and paragraphs. These models functioned by utilizing memory of previous inputs. Although introduced in the 1980s, RNNs gained renewed interest in the 2010s with
the emergence of long short-term memory (LSTM) models, which addressed the vanishing gradient problem and facilitated modeling long-term text dependencies \cite{hochreiter_long_1997}. However, LSTMs still
faced issues during training, such as vanishing gradients and non-parallelizable operations due to their sequential nature. Attention mechanisms addressed these challenges by allowing models to selectively attend to relevant parts of a sequence, thus improving language modeling, machine translation, and more, while also creating a way to parallelize the training \cite{vaswani_attention_2017}. Shortly after their introduction, transformer-based models like BERT \cite{devlin_bert_2019} and GPT achieved state-of-the-art results on various NLP tasks and enabled new applications, such as large-scale language generation. As a case in point, the development of GPT models led to the
creation of ChatGPT, an example of a transformer-based LLM \cite{radford_learning_2021}. These pre-trained, transformer-based models have demonstrated strong performance across a wide range of NLP tasks in the
field of education and beyond.

\subsubsection{Large Language Models }\label{large-language-models}

Large language models are a class of NLP models that harness vast amounts of data and computational power to learn intricate language patterns and (for some models) generate contextually relevant responses \cite{brown_language_2020}. Typically pre-trained on extensive text corpora (e.g., a common crawl of the web consisting of billions of tokens of text) and fine-tuned for tasks such as question-answering or text classification \cite{devlin_bert_2019}, LLMs are distinguished by their
capacity to seemingly comprehend and generate human-like text, rendering them eligible for application in a broad array of NLP tasks and contexts \cite{radford_learning_2021}. The advancement of LLMs has been propelled by progress in deep learning techniques, particularly the development of transformers, a type of neural network architecture designed for NLP tasks through attention mechanisms \cite{vaswani_attention_2017}. These mechanisms enable the model to concentrate on relevant text segments and presented a significant improvement over previous state-of-the-art LSTM models as mentioned above. Notable LLMs include GPT-3, developed by OpenAI, which exhibits impressive language understanding and generation capabilities \cite{brown_language_2020}. Prior to GPT-3, at the time it was released, Google's BERT set new benchmarks for various NLP tasks, such as sentiment analysis and question-answering \cite{devlin_bert_2019}. Other prominent LLMs include AllenAI's ELMo, which computes contextualized word representations to enhance performance on diverse NLP tasks \cite{peters_deep_2018}; Facebook's RoBERTa, a revised version of BERT that achieves state-of-the-art results on multiple NLP benchmarks \cite{liu_roberta_2019}; and XLNet, a Google/CMU collaboration that applies transformer models to generalize BERT while improving computational efficiency \cite{yang_xlnet_2019}.

Recent LLMs employ self-supervised learning algorithms on larger datasets to foster more general language understanding. For instance, Language Models are Unsupervised Multitask Learners (LM-USTL) applies
Constitutional AI for language model alignment, yielding strong results on unsupervised NLP tasks \cite{radford_language_2019}. Other leading self-supervised LLMs include OpenAI's GPT-3 \cite{brown_language_2020}, Google's T5 \cite{raffel_exploring_2020}, and Anthropic's Claude \cite{wallace_universal_2021}. These generalist LLMs can be fine-tuned on specific downstream tasks to enhance performance. State-of-the-art LLMs such as GPT-3, BERT, and
XLNet have achieved human-level performance on numerous NLP tasks. Continued advances in NLP are expected to enable LLMs to attain deeper, more nuanced language understanding, rivaling or even surpassing human capabilities in some use cases. Such anticipated progress points to opportunities to use LLMs in support of educational research.

\subsection{Applications of NLP and LLMs in Education
Research}

By partially automating the text analysis process, NLP techniques can help educators and researchers assess students' understanding of complex concepts, evaluate their argumentation skills, and identify misconceptions in their written work \cite{graesser_coh-metrix_2011,shermis_contrasting_2015}. Additionally, LLMs can be employed to provide feedback on students' writing, enhancing the learning experience and supporting the development of writing skills \cite{dikli_overview_2006,roscoe_shallow_2013}. Beyond analyzing student writing, NLP and LLMs can also be utilized to assess learning outcomes in educational settings. For instance, these techniques have been applied to analyze students responses to open-ended questions, facilitating the measurement of their understanding and progress \cite{ha_applying_2011}. Furthermore, NLP tools can
be used to evaluate students' performance in online discussions and collaborative learning environments, providing insights into their engagement, critical thinking, and knowledge construction \cite{wise_learning_2017}.

In the following review on the use of NLP in education, we discuss studies in which the basic mechanism for assessing students' descriptive answers involved comparing them with reference or graded answers. Early research, such as \cite{jordan_assessment_2009} and \cite{butcher_comparison_2010}, employed rudimentary NLP approaches, such as keyword matching for this purpose. However, these methods were limited as they lacked knowledge of grammar or syntax, and were prone to failures due to different words with the same meaning or many possible meanings for a word or phrase. Subsequently, more sophisticated NLP techniques based on pattern matching emerged, with n-grams being a common model. Examples include \cite{nehm_transforming_2012} and \cite{haudek_what_2012} in the field of
biology. \citet{galhardi_machine_2018} who reviewed 44 papers on ML methods in ASAG systems. They found that n-grams were the most common model, used in over 70\% of the reviewed studies. Approaches like Bag-of-Words (BoW) and Term Frequency-Inverse Document Frequency (TF-IDF) were employed to create crude vector representations of documents by simply counting the instances of each word in the document (BoW model) and then down weighting words based on how frequently they appear in the entire corpus of text (TF-IDF model). While functional, these approaches discard a lot of useful information encoded in the text.

Researchers have also integrated NLP feature extraction techniques with supervised or unsupervised ML models, such as Support Vector Machines (SVM), Decision Trees, and k-nearest neighbor algorithms, to automate the grading of students' answers \cite{blessing_machine_2021}. For example, \citet{nehm_transforming_2012}
developed the Summarization Integrated Development Environment (SIDE) tool, which used BoW and SVM for classification. \citet{moharreri_evograder_2014} and 
and \citet{yik_development_2021} applied similar techniques for evaluating students' written responses in evolutionary biology. \citet{lee_automated_2019} built automated scoring models using c-rater-ML, an NLP engine developed by Educational Testing Service, which extracts several types of textual features and employs support vector regression methods.

Ensemble learning, which trains multiple models and combines them to make predictions, has been combined with the BOW approach for better generalization performance. \citet{roy_iterative_2016} proposed a computer assessment tool based on an ensemble of two classifiers, capturing complementary information useful for grading student answers. \citet{jescovitch_comparison_2021} trained a  model using an 8-class classification algorithm ensemble, treating the task of assigning scores to student writing as a text classification problem.

In recent years, NLP lexical feature extraction has been combined with neural network algorithms. For example, \citet{zhang_natural_2022} and \citet{zhai_applying_2022} demonstrated that RNNs can extract latent features more representative than the original input features. In another example, \citet{maestrales_using_2021}, who used
human-scored responses to train machines and develop algorithmic models by examining lexical features like n-grams. Likewise, \citet{vairinhos_framework_2022} introduced a text-mining-based framework designed to grade open-ended questions by extracting relevant textual features. In addition to this kind of text feature extraction, there are also popular sources and databases that researchers use such as WordNet. As one of
the most widely used knowledge-based sources, WordNet calculates the similarity between two words by leveraging a large lexical database of English words and their relations. Many studies have utilized WordNet
and knowledge-based features to analyze student responses, such as \cite{pribadi_automatic_2017, shaukat_semantic_2021}. These methods involved
calculating similarity matrices and utilizing various similarity measures, including the Dice Coefficient and block distance similarity measure.

Corpus-based similarity, on the other hand, constructs a knowledge space using large corpora, representing words as vectors in a multi-dimensional semantic space. Examples already mentioned above in the abbreviated overview of NLP include Word2Vec and Glove \cite{mikolov_efficient_2013,pennington_glove_2014}. As mentioned in the prior section, embedding models such as Word2Vec and Glove help create a new set of features that researchers can use for short answer grading. Several studies have utilized such an approach \cite{guerrero_using_2019,forsyth_short_2021,romero-gomez_natural_2022}. Embedding models help lead into transformer-based models by forming some of the initial input to those models.

Building on top of modern, popular neural network architectures, transformer-based models like BERT have demonstrated improved performance for analyzing student writing in various contexts. For example, \citet{liu_automated_2022} reported substantial improvement in cognitive engagement recognition using a convolutional BERT-CNN. Likewise, \citet{wulff_bridging_2022,wulff_enhancing_2023} showcased the potential of unsupervised ML models combined with pre-trained language models like BERT for exploring themes in physics teachers' written responses. Their data contained reflection responses of pre-service physics teachers while watching a video vignette. \citet{wulff_bridging_2022} concluded that their NLP workflow was successful in classifying higher level reasoning sentences in the written reflections according to the reflection-supporting theoretical model. In another study using BERT as
the backbone for the NLP model, \citet{sung_improving_2019} investigated the enhancement of pre-trained contextual representations for ASAG and reported up to 10\% absolute improvement in macro-average-F1 on benchmark datasets. \citet{camus_investigating_2020} also demonstrated how fine-tuning different pre-trained Transformer-based
architectures, could results in up to 13\% absolute improvement in macro-average-F1 over state-of-the-art results at the time. Finally, in another very recent study, \citet{wulff_enhancing_2023} explored machine
learning (ML) and NLP techniques for enhancing writing analytics in science education research. By using BERT in conjunction with the dimension reduction technique UMAP and the clustering method HDBSCAN, the researchers demonstrated the potential of ML and NLP for formatively assessing preservice teachers' written reflections. Results like these highlight the utility of transformer-based models in
analyzing student writing. Our work extends this body of research on applications of transformer-based LLMs for educational research.

\subsection{Significance of the Current Study}

The current study aims to address the limitations of previous research on NLP and LLMs to support education research by pairing existing methods with novel generative text models. The goal is to broaden the community's knowledge of how these models function and encourage the
models' careful incorporation into engineering education research because the models have a better ability to represent semantic meaning than prior iterations. By exploring the application of LLMs in the analysis of engineering students' career interests essays, this study contributes to the understanding of how these advanced models can be used to gain valuable insights into student
learning and development. Specifically, we aim to demonstrate a set of methods for utilizing NLP and LLMs to thematically analyze unstructured text data in engineering education. This is only a subset of methods that can be adapted and employed in various research contexts and
educational settings, fostering further innovation and exploration of NLP and LLMs in engineering education research. Other use cases could extend beyond student writing to other populations (e.g., instructors,
working professionals), different kinds of data (e.g., open-response questions on surveys, interview transcripts), and broader research questions.

\section{Methods}\label{methods}

We used a combination of embedding models and generative text models in this study to illustrate ways to leverage LLMs in engineering education research. The goal was to be able to identify themes in student writing in an accurate and scalable manner using either a deductive (pre-determined codes) or inductive (creating a codebook from scratch) approach. In some ways, this mimics a thematic analysis \cite{clarke_thematic_2017} approach to  ualitative data analysis insofar as the eventual product may look similar (thematically labeled texts). In this Methods section, we explain how to leverage two key advancements in modern NLP research to accomplish the analysis task: text embeddings and transformer-based LLMs. The embeddings enable us to cluster semantically similar text together to assist with codebook generation while the generative LLMs allow us to attempt to create those codes and then apply the codes to unlabeled text. Then, through prompt engineering, we were also able to instruct the generative text model to self-evaluate and rate the accuracy of the labels it applied in a previous step. In the following subsections, we discuss the data used for evaluation, data pre-processing, how each step in the process works, and our evaluation approach. Code and notebooks for completing these steps will be available at {[}redacted GitHub repo{]}. (NB: The words code and label are overloaded here and also sometimes used interchangeably. The word `code' can refer to the actual script written in the python programming language or a short descriptive phrase to describe a span of text. Likewise, the word `label' can be used as a verb, as in ``label the text with a code'' or as a noun and therefore synonymously with code, as in ``apply a label to the
text'').~

\subsection{Data Collection}\label{data-collection}

Since the purpose of this paper was to demonstrate options for thematically labeling text in student writing in engineering education research, for this study we used essays written by first-year engineering students. The essays were part of an assignment asking them to describe their career interests and to explore career options. The specific assignment prompt stated:~

\emph{The first step in developing your portfolio is to identify a potential career goal and relevant job. To be useful, your goal should be concrete and include not only the engineering field you're interested in, but the kind of work you'd like to do -- e.g. ``working as an electrical engineer for a power company'' or ``designing safety equipment for mining operations.''}

\emph{In addition to defining a goal, you also need to select a specific job you would be interested in. Note that choosing a goal and choosing a job might be iterative; a general goal might lead you to some specific job ideas, which might in turn help you make your goal more concrete. In addition to identifying and describing your goals, you also need to explain why you selected these goals -- how do they match your interests, your strengths, your aspirations, etc.}

Essays typically ranged in length from 500 to 1,500 words. Data used for this study were from a subset of sections in the 2018 and 2019 administrations of this course. In total, we used 1,014 of these essays for this study.~

\subsection{Data Pre-processing}\label{data-pre-processing}

To analyze these data, we first converted all of the assignment .pdf files into .txt files. Next, we removed parenthetical references, identifying information (i.e., student names), and other extraneous formatting (e.g., headings). We then parsed the essay text by splitting essays at the sentence level. We did this in order to capture themes and topics at a higher resolution --- in some ways, labeling themes in sentences can tell more about what a student is discussing than working at the whole paragraph or whole essay levels. Ideally there could be a middle ground wherein one could identify sentence fragments or multi-sentence segments for this purpose, but to our knowledge that level of flexible parsing is more challenging at this time. As the context window for generative models continues to grow, one might anticipate being able to skip this step entirely for some use cases. Following this text cleaning process, the dataset consisted of a collection of sentences for each student. In this study, that translated to 9,105 sentences from 1,014 essays.

\subsection{Data Analysis}\label{data-analysis}

\begin{figure}[htbp]
\centering
\includegraphics[width=0.8\linewidth]{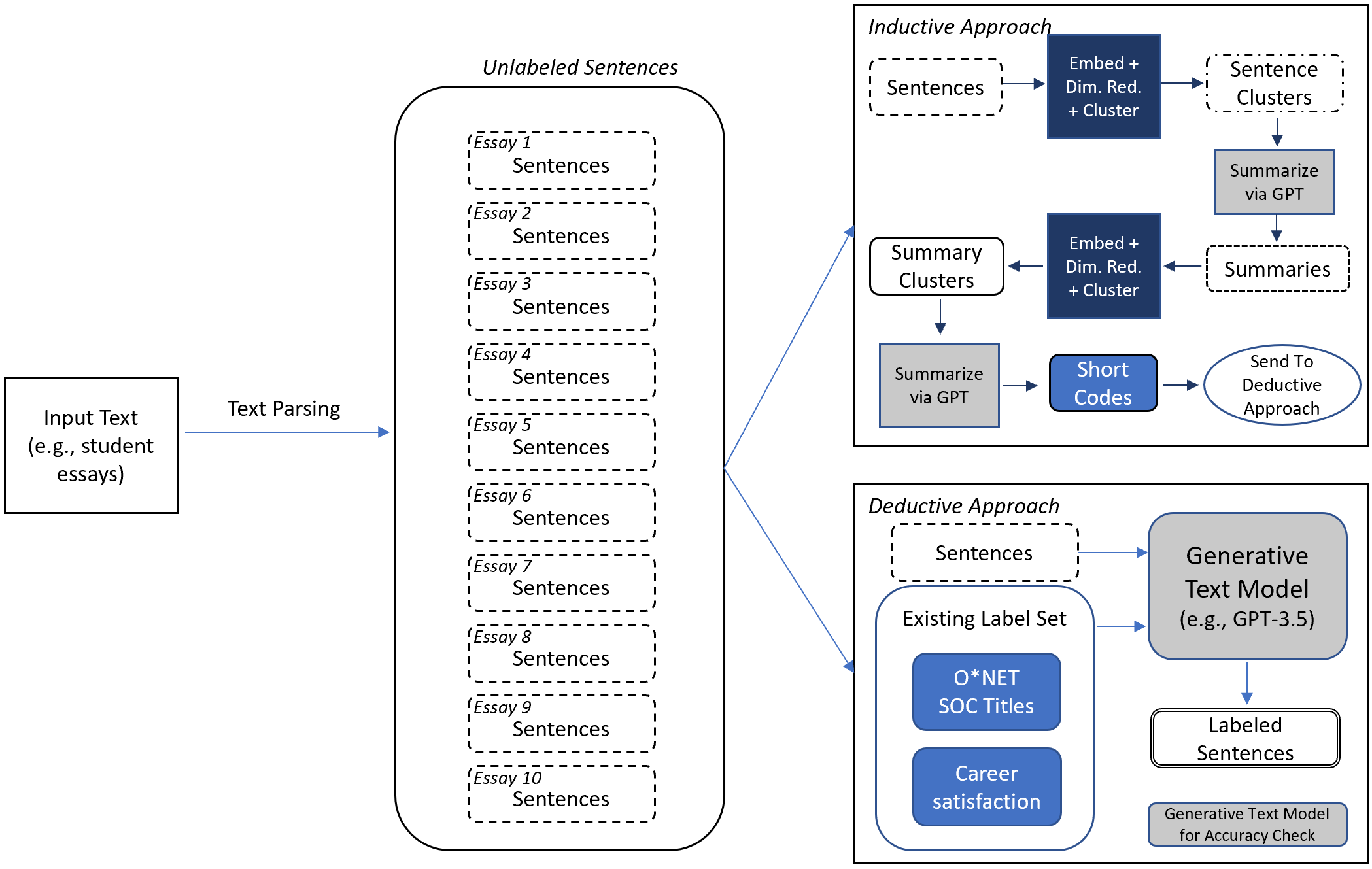}
\caption[Inductive and Deductive Approaches to Analyzing Student Essays]{Inductive and Deductive Approaches to Analyzing Student Essays\protect\footnotemark}\label{fig-nlp-workflow}
\end{figure}

To analyze the student essays, we used two approaches -- an inductive approach and a deductive approach. The inductive approach involved using a combination of a text embedding model and clustering of those text embeddings to identify semantically similar statements from students -- e.g., ``I want to help protect the earth's resources'' and ``I would like to work toward environmental sustainability in my career''. We then used the generative text model to label each semantically similar group of statements and used that label as the code for the codebook. In the deductive approach, we started directly from a list of labels/codes derived from a different mechanism. In this study, those labels were from (a) Occupational Network (O*NET) Standard Occupation Code (SOC) job titles or (b) options on an existing survey about career satisfaction. The general workflow is shown in figure \ref{fig-nlp-workflow}.

\footnotetext{This flowchart illustrates the main steps taken in both the inductive and deductive approaches to analyzing student essays using natural language processing techniques. In the inductive approach, text embeddings and clustering are used to generate and refine themes, while in the deductive approach, pre-existing labels are applied to label the text.}

\subsubsection{Inductive Approach for Theme
Identification}\label{inductive-approach-for-theme-identification}

To identify themes in student essays, we used a four-step process. Those steps are outlined below:

\textbf{Step one: Embed and cluster
sentences}

To analyze the topics mentioned in  students' written responses, we first embedded each sentence into a continuous vector space using a pre-trained transformer encoder model. There are many options to do this, including those from the HuggingFace model repository \cite{wolf2019huggingface}. For this study, we used a Sentence Transformers embedding model because it is open source and benchmark test results suggest the model maintains high performance across a variety of tasks \cite{reimers-2019-sentence-bert}. We specifically used the mpnet model due to its higher overall performance on benchmark tests compared to other similar models. The key idea behind an embedding model is to generate a high-dimensional numeric representation of each unit of text. In this study, those units were sentences. With this numeric representation for each sentence, one can then perform mathematical operations such as dimension reduction or clustering, both of which we used here. Theoretically, at the end of the process one will have clusters of semantically similar text.

We clustered these embeddings of student essay sentences using agglomerative clustering. These initial clusters contained sentences on a common theme or subject matter. The goal of this step was to group statements such as ``My goal is to work as a cybersecurity specialist'' and ``I want to protect against cyberattacks'' since they are semantically similar. The challenge in that task is that statements will often share semantic similarity without sharing syntactic or lexical similarity---that is, there are many ways to communicate the same idea. Modern, transformer-based LLMs typically offer the ability to capture this semantic meaning well, which is one of the reasons we believe these models present opportunities for education researchers to scale qualitative data analysis.

\textbf{Step two: Generate 15-word summaries of those clusters}

With the original set of sentences clustered together, we then proceeded cluster-by-cluster and provided the sentences for each cluster to a generative language model, i.e., GPT-3.5, along with a prompt to generate a 15-word summary of the sentences in that cluster. For the example in the preceding paragraph, ideally the summary would identify `cybersecurity' as the main theme. These summaries were intended to distill the essence of each cluster into a short phrase. The specific prompt we used for this step was: ``The following comments were written in response to the question 'Please describe your future career interests.'. What do they have in common? Keep your response under 15 words. Start your response with `Commonality: '. \textbackslash{}n COMMENTS:''.

\textbf{Step three: Cluster the
summaries}

Creating clusters of statements and then generating
descriptive summaries of each cluster could still lead to much overlap between clusters. To address this issue of overlapping clusters, we used a second round of clustering. With this second round, instead of clustering students' written sentences, we clustered the model-generated summaries of those sentence clusters. This involved embedding the 15-word summaries and clustering those embeddings using the same process as with the original student statements. For example, the process grouped together first-round summaries such as:

\begin{itemize}
    \item ``Interest in environmental conservation, specifically in reducing plastic pollution and protecting marine life''
    \item ``Interest in careers focused on sustainability, renewable energy, and environmental protection''
    \item ``Interest in environmental sustainability and addressing climate change''.
\end{itemize}

The summary clusters from this step represented the major themes discussed across student responses.

\textbf{Step four: Generate short labels of those summary groupings}

Finally, with the sentence summaries clustered, we again prompted GPT-3.5 to generate a three-to five-word label for each of the summary clusters to capture the theme they represented. In the example provided in step three, the label could be ``environmental sustainability careers''.

Overall, this multi-step process can leverage the capabilities of a modern generative language model to analyze open-ended student writing and extract the major topics of discussion at increasing levels of abstraction and conciseness. The resulting set of labels then represents most of the key themes that could be gleaned from the student writings using this data-driven approach. The model-generated labels also create an initial codebook, with the initial round of summaries representing sub-codes and the final round of summaries representing parent codes. In that way, the process resembles a bottom-up codebook generation technique. These labels, and the summaries that produce them, might provide useful insights beyond what manual analysis alone can achieve because of the ability to scale the level of analysis. However, if one were approaching their research and analysis with an extant theory, then a deductive labeling approach may be more appropriate. We describe such an option next.

\subsubsection{Using a Deductive Labeling
Approach}\label{using-a-deductive-labeling-approach}

In many situations a researcher may already know the topics they are looking for. In those instances, they can use a deductive approach in which they simply start with that label set, use it in a prompt to the generative model, and receive a response back from the model based on that prompt. In the case of student career interest essays, one might want to know something simple such as (a) the engineering field or career path each student mentions or (b) factors students identify as important to their career satisfaction. Using an older NLP approach, the researcher could then generate a potential list of words (e.g., environmental, environment, sustainability, renewable) and look specifically for those words; however, that dictionary creation adds an unnecessary layer of work and can still be brittle if the list is not completely exhaustive. Instead, leveraging the LLM's ability to capture semantic meaning could be a more feasible approach. The question then becomes, ``where should my set of labels come from?'' For this study, we demonstrate the deductive approach with a set of (a) job titles from the US O*NET SOC codes and (b) factors associated with career satisfaction derived from an existing survey of engineering student career interests. We describe each label set in the following subsections.

\hypertarget{onet-soc-labels}{%
\paragraph*{\texorpdfstring{O*NET SOC Labels
}{O*NET SOC Labels }}\label{onet-soc-labels}}
\addcontentsline{toc}{paragraph}{O*NET SOC Labels }

The first set of labels we used for the deductive approach is for a scenario wherein a researcher is interested in identifying which careers or career sectors students wanted to pursue. Although one could simply give students a list to choose from, asking them to express their interests in their own words can also provide that information but with less priming to answer a particular way. To accomplish the goal of understanding which jobs they may want to pursue as engineers, we used the O*NET SOC codes and job title for a candidate list. We only used the titles for levels 15 (Computer and Mathematical Occupations) and 17 (Architecture and Engineering Occupations) to balance the tradeoff between model memory and a comprehensive list. The list of 79 titles we used for this demonstration is in table \ref{tab-onet-titles}. Undoubtedly there will be students interested in careers outside this list, which is a limitation of this approach.

\begin{table}[htbp]
\caption{O*NET Job Titles}\label{tab-onet-titles}
\begin{tabular}{@{}p{0.3\linewidth} p{0.65\linewidth}@{}}
\toprule
Category & Occupation \\
\midrule
Computer and Mathematical Occupations & 'Computer Systems Analysts', 'Health Informatics Specialists', 'Information Security Analysts', 'Computer and Information Research Scientists', 'Computer Network Support Specialists', 'Computer User Support Specialists', 'Computer Network Architects', 'Telecommunications Engineering Specialists', 'Database Administrators', 'Database Architects', 'Data Warehousing Specialists', 'Network and Computer Systems Administrators', 'Computer Programmers', 'Software Developers', 'Software Quality Assurance Analysts and Testers', 'Web Developers', 'Web and Digital Interface Designers', 'Video Game Designers', 'Web Administrators', 'Document Management Specialists', 'Information Security Engineers', 'Digital Forensics Analysts', 'Blockchain Engineers', 'Computer Systems Engineers/Architects', 'Information Technology Project Managers', 'Actuaries', 'Mathematicians', 'Operations Research Analysts', 'Statisticians', 'Biostatisticians', 'Data Scientists', 'Business Intelligence Analysts', 'Clinical Data Managers', 'Bioinformatics Technicians' \\
\midrule
Architecture and Engineering Occupations & 'Architects', 'Landscape Architects', 'Cartographers and Photogrammetrists', 'Surveyors', 'Geodetic Surveyors', 'Aerospace Engineers', 'Agricultural Engineers', 'Bioengineers and Biomedical Engineers', 'Chemical Engineers', 'Civil Engineers', 'Transportation Engineers', 'Water/Wastewater Engineers', 'Computer Hardware Engineers', 'Electrical Engineers', 'Electronics Engineers, Except Computer', 'Radio Frequency Identification Device Specialists', 'Environmental Engineers', 'Health and Safety Engineers, Except Mining Safety Engineers and Inspectors', 'Fire-Prevention and Protection Engineers', 'Industrial Engineers', 'Human Factors Engineers and Ergonomists', 'Validation Engineers', 'Manufacturing Engineers', 'Marine Engineers and Naval Architects', 'Materials Engineers', 'Mechanical Engineers', 'Fuel Cell Engineers', 'Automotive Engineers', 'Mining and Geological Engineers, Including Mining Safety Engineers', 'Nuclear Engineers', 'Petroleum Engineers', 'Energy Engineers, Except Wind and Solar', 'Mechatronics Engineers', 'Microsystems Engineers', 'Photonics Engineers', 'Robotics Engineers', 'Nanosystems Engineers', 'Wind Energy Engineers', 'Solar Energy Systems Engineers' \\
\botrule
\end{tabular}
\end{table}

To apply these labels, we instructed the generative model with the prompt listed in appendix section \ref{app-prompt-onet}.

\hypertarget{career-satisfaction-labels}{%
\paragraph*{Career Satisfaction
Labels}\label{career-satisfaction-labels}}
\addcontentsline{toc}{paragraph}{Career Satisfaction Labels}

Rather than knowing about careers that students want to pursue, one could also study the importance of various factors to student anticipated career satisfaction. For this demonstration, we used a list of items from a prior survey given to engineering students \cite{shealy2017survey}. The survey contained a series of questions about the importance of various opportunities and characteristics to their anticipated career satisfaction. Table \ref{tab-car-sat-facts} provides the list of factors.

\begin{table}[htbp]
\caption{Career Satisfaction Factors}\label{tab-car-sat-facts}
\begin{tabular}{@{}ll@{}}
\toprule
 \\
\midrule
Making money & Having lots of personal and family time\\
Becoming well known & Having an easy job\\
Helping others & Being in an exciting environment\\
Supervising others & Solving societal problems\\
Having job security and opportunities & Making use of my talents and abilities\\
Working with people & Doing hands-on work\\
Inventing/designing things & Applying math and science\\
Developing new knowledge and skills & Volunteering with charity groups\\
\botrule
\end{tabular}
\end{table}

To apply these labels, we instructed the generative model with the prompt listed in appendix section \ref{app-car-sat-prompt}.

\subsubsection{Accuracy Checks}\label{accuracy-checks}

We also tested the model's ability to check its own accuracy. Such an ability is important to ensure higher accuracy rates and avoid the need for a human to manually check \emph{all} of the results, which would nullify some of the advantages gained by leveraging LLMs to do this work. If a model could accurately flag instances that required a second check from a person then that would still present an improvement and reduction in time. To accomplish this accuracy check, we prompted the same GPT-3.5 model with the label that had been previously applied and instructions to rate the accuracy of that label on a scale from 1 (completely inaccurate) to 10 (completely accurate). We then checked that rating with a human's rating of the same label for the same comment. We noted where there was agreement or disagreement with the accuracy check model. For example, if a comment was originally labeled as ``environmental engineer'', the accuracy check gave it a high (\textgreater{} 6) accuracy rating. If the comment was indeed about environmental engineering, then that was agreement. The converse involved a scenario where a label was applied, the model rated it highly, but in fact that was an inaccurate label as judged by the human evaluator (this would be an instance of disagreement). The goal here was to test whether a prompted generative model could act as its own accuracy check. An example prompt for the career satisfaction labels is the following:

``On a scale from 1 (completely inaccurate) to 10 (completely accurate), rate the accuracy of the following topic label for describing one of the things a student said about factors important for their career interests. Only provide your numeric rating.

LABEL: \textless Insert label\textgreater.

COMMENT: \textless Insert comment\textgreater.''

An example of the returned response and the human rater's assessment of the model's accuracy score is given in table \ref{mod-acc-ck-ex}.

\begin{table}[htbp]
\caption{Example model accuracy check}\label{mod-acc-ck-ex}
\begin{tabular}{@{}p{0.35\linewidth}lll@{}}
\toprule
Student comment & Model label & Model accuracy score & Humar rater eval \\
\midrule
Also, I'd love to reach that goal because the jobs at NASA for engineers pay well and one thing I want to be sure about in my future is to be financially stable. & Making money & 8 & Agreement\\
\botrule
\end{tabular}
\end{table}

\section{Results}\label{results}

In the following sections we review the results of applying the inductive and deductive coding approaches with LLMs to analyze 1,014 student essays. The first section presents the inductive approach for identifying the topics students discussed in their essays. The second section presents the deductive method to apply a priori codes to identify whether a student was discussing specific topics in their essay. As a reminder, the topics used here were either specific careers or factors important to their career satisfaction.

\subsection{Inductive Approach: Clustering and Summarizing to Analyze
Essays}\label{inductive-approach-clustering-and-summarizing-to-analyze-essays}

We first parsed the 1,014 essays into their 9,105 constituent sentences and then clustered those sentences into 285 semantically similar groups. A researcher could read each cluster or prompt a language model to summarize them. As mentioned in the Methods section, we chose the latter. We will refer to these as sentence-cluster summaries. Table \ref{tab-ex-env-issues} shows examples of sentence clusters and their model-generated sentence-cluster summaries for statements about environmental issues.

\begin{table}[!htbp]
\caption{Examples of environmental issues clusters}\label{tab-ex-env-issues}%
\begin{tabular}{@{}p{0.65\linewidth} p{0.3\linewidth}@{}}
\toprule
Student statements in cluster* & Model-generated Summary of statements in cluster (sentence cluster summaries) \\
\midrule
Having access to clean water is a huge problem in the world. &
\multirow{4}{0.99\linewidth}{Interest in addressing global issues related to access to clean water and affordable healthcare.}\\
One would think that because water is so important for life, it is
prioritized with the creation of new technology and at the forefront of
political talk.\\
Many people die from diseases such as cholera and simply from lack of
water.\\
Also, many people did not have clean water and sewage systems, which
could lead to the transfer of many diseases. \\
\midrule
Being able to work on restoring the environment would be very
rewarding. & \multirow{4}{0.99\linewidth}{Environmental and humanitarian interests, desire to use
engineering to solve global problems.}\\
I would be enticed to find cheaper methods of rehabilitation so that the
same amount of money would be able to restore the land more
thoroughly.\\
I am interested in cleaning the ocean because I want to make sure my
children, grandchildren, and great grandchildren have the opportunity to
go scuba diving in a clean and habitated ocean.\\
I am hoping that I can also work with things like water conservation and
maybe even some involvement with water canals and drainage
systems. \\
\midrule
In high school I enjoyed studying the environment and sustainability. & \multirow{4}{0.99\linewidth}{Interest in environmental sustainability and addressing climate change.}\\
I have had a deep passion for the environment and conservation from a
young age and I strive to be a part of the solution rather than the
problem.\\
The main reason I want to help save the earth is so that not only I can
live a life without worrying how much longer our planet can support us,
but so generations after me can live this way as well.\\
Motivation is very important for my work, and I can envision myself
waking up every day motivated to save the Earth.  \\
\botrule
\end{tabular}
\footnotesize{* Some statements were removed from each cluster in this table for space purposes}
\end{table}

\FloatBarrier

For a second demonstration, table \ref{tab-ex-cyb-iss} shows examples of sentence clusters and their model-generated sentence-cluster summaries for statements  about careers in cybersecurity.

\begin{table}[!htbp]
\caption{Examples of cybersecurity clusters}\label{tab-ex-cyb-iss}
\begin{tabular}{@{}p{0.65\linewidth} p{0.3\linewidth}@{}}
\toprule
Student statements in cluster* & Model-generated Summary of statements in cluster \\
\midrule
As a cyber security analyst, I would be able to help protect people from the ever-growing amount of threats that are targeting the technology around us. & \multirow{5}{0.99\linewidth}{Interest in cyber security, protecting data and addressing security flaws and threats.}\\
I am fascinated to learn how our adversaries attempt to break into our systems through exploiting the software code and how our country defends against it as well as going on the offense to break into others. \\
For me, I want to be able to protect people's information from malicious intent. \\
Problems that I would address are security flaws in all kinds of electronic devices and sites. \\
The work I would be doing is protecting people from malicious software and other cyber security threats towards their data or privacy. \\
\midrule
Till know I don't know what I am going to provide to the customers, but I am pretty sure it is going to be related with AI and cyber security. & \multirow{4}{0.99\linewidth}{Interest in pursuing a career in cybersecurity and/or computer science with a focus on AI and security.} \\
There is tons of work in Cybersecurity, so learning encryption is and securing data is always something I could work towards, and A.I. is also a booming research field in the software and computer science world, so I could definitely go work in anything that might benefit from A.I. concepts such as the self-driving car that people want so much to minimize human accidents on the road and save many lives.\\
I want to work in cybersecurity because I like coding, as well as decoding, and I want to explore ways to keep others and myself safe in our use of technology. \\
Currently, I want to be a cyber security analyst because of the satisfaction I get from dealing with cyber issues and the rush associated with fighting malicious people on the internet. \\
\midrule
My goal for the future is to either work as a software developer at whatever company is working on creative, innovative software or to work for a government agency such as the FBI or NSA as a cyber security analyst. & \multirow{6}{0.99\linewidth}{Interest in cybersecurity and government agencies/companies in the field.} \\ 
A specific career goal that I would want to reach someday is working for the Central Intelligence Agency as a cyber security officer. \\
I would like to be involved with cybersecurity, network infrastructure, or big data in the computer science field through a job with the government. \\
My current career goal as of right now is to be an intelligence analyst at the NSA or CIA. \\
I hope to one day work for the Central Intelligence Agency with the job of researching malware and developing tools to defend against it. \\
I want to work for the CIA as a Cyber Threat analyst in the cybersecurity division. \\
\botrule
\end{tabular}
\footnotesize{* Some statements were removed from each cluster in this table for space purposes}
\end{table}

Next, with the 285 sentence-cluster summary statements, we clustered \emph{those} statements to identify a smaller set of themes since several clusters shared overlapping themes (e.g., more than one cluster of sentences was about working toward environmental sustainability, as shown in table \ref{tab-ex-env-issues}). The second round of clustering generated 36 clusters. This process helped distill the information in separate clusters. Such a distillation makes it easier to identify the group of statements about working for environmental causes and a separate group of statements about working in the cybersecurity field. With these 36 clusters, we again prompted the generative text model to summarize those clusters of summaries. We call these second-round summaries or summary-cluster summaries. Examples of these results for five of the 36 clusters are presented in table \ref{tab-2-rd-summ}. The five clusters are labeled ``Interest in environmental issues'', ``Passion for cars and engineering'', ``Interest in computer science/engineering and software development careers'', ``Interest in problem-solving and innovation'', and ``Desire to make a difference''.

\begin{longtable}{@{}p{0.65\linewidth} p{0.3\linewidth}@{}}
\caption{Examples of clusters of sentence-summary clusters and model-suggested labels for the summary clusters}\label{tab-2-rd-summ} \\
\toprule
Model-generated summary of statements in sentence-cluster & Model suggestion (summary-cluster summaries)\\
\midrule
\endfirsthead
\multicolumn{2}{c}%
{{\bfseries Table \thetable\ continued from previous page}} \\
\toprule
Model-generated summary of statements in sentence-cluster & Model suggestion (summary-cluster summaries)\\
\midrule
\endhead
\bottomrule
\multicolumn{2}{r}{{Continued on next page}} \\
\endfoot
\bottomrule
\endlastfoot
Interest in environmental conservation, specifically in reducing plastic pollution and protecting marine life.  & \multirow{15}{=}{Interest in
environmental issues} \\
Interest in careers focused on sustainability, renewable energy, and environmental protection. \\
Interest in nuclear fusion and addressing related problems in research and engineering. \\
Interest in addressing global issues related to access to clean water and affordable healthcare. \\
Environmental and humanitarian interests, desire to use engineering to solve global problems. \\
Interest in environmental sustainability and addressing climate change. \\
Interest in renewable/sustainable energy and improving energy
efficiency. \\
Interest in renewable energy and sustainability, including solar and
wind power. \\
Interest in renewable/clean energy and addressing climate change. \\
Interest in addressing environmental issues, particularly related to energy production and reducing carbon emissions. \\
Interest in addressing environmental issues, particularly water pollution and sustainability. \\
Interest in using chemical engineering to address environmental and sustainability issues, including alternative energy sources. \\
Interest in environmental engineering, sustainability, and making a positive impact on the environment. \\
Interest in addressing environmental issues in mining and promoting sustainability. \\
Interest in alternative and sustainable energy sources, particularly nuclear fusion. Concerns about environmental impact and negative stigma. \\
Interest in renewable energy and related industries, such as
sustainability and nuclear energy. \\
Environment and sustainability concerns.\\
\midrule
Interest and passion for coding and computer science as a career choice. & \multirow{11}{=}{Interest in computer science/engineering and software development careers}\\
Interest in pursuing a career in computer science or computer
engineering. \\
Future career interests involve software development, improving skills,
and working with technology. \\
Interest in working as a software engineer/developer for Apple,
specifically on operating systems and app development. \\
Interest in learning and mastering multiple programming languages for
career success. \\
Interest in pursuing a career in computer science or computer
engineering, often sparked by taking programming classes in high
school. \\
Interest in computer science, programming, and software development as a
career path. \\
Future career interests in computer science/engineering, job security,
variety of job opportunities, good pay and benefits. \\
Career interest in software development/engineering, with goals of
working for major companies or starting own firm. \\
Future career interests involve software engineering, programming, and
problem-solving. \\
Future career interests involve computer science, programming, math, and
potential innovation/advancement in the field.  \\
\midrule
Passion for cars and hands-on experience with cars, often influenced by family members. & \multirow{6}{0.99\linewidth}{Passion for cars and engineering} \\
Interest in working as a mechanical engineer in the automotive industry,
specifically for well-known companies or race teams. \\
Interest in automotive engineering, designing and testing car parts,
improving safety and efficiency. \\
Interest in working with cars and pursuing a career in the automotive
industry. \\
Interest in working for car manufacturers, particularly BMW, Mercedes,
Porsche, Ferrari, and Toyota. \\
Interest in cars and their mechanics, particularly designing and
improving various components. \\
Interest in cars, speed, and technology. \\
Interest in automotive engineering/design, specifically in creating more
efficient and powerful engines and vehicles.  \\
\midrule
Interest in innovation, technology, and problem-solving for future advancements. & \multirow{13}{=}{Interest in problem-solving and innovation} \\
Working in a team to design and develop products or technology, using CAD and problem-solving skills. \\
Emphasis on testing, programming, problem-solving, and design. \\ Preference for hands-on, active work outside of a traditional office setting. \\
Interest in hands-on design and creation processes, curiosity about how things work. \\
Interest in problem-solving, logical thinking, and spatial awareness. \\
Interest in engineering and problem-solving across various industries and specializations. \\
Interest in computer programming and its connection to mathematics and problem-solving. \\
Interest in hands-on problem solving, creative thinking, and finding efficient solutions. \\
Interest in hands-on work, design, testing, and innovation in a project-oriented job. \\
Interest in project-based work, design, testing, and monitoring/evaluation in various fields. \\
Problem-solving is a key skill required in future career interests. \\
\midrule
Interest in healthcare industry, helping others, personal/familial experiences, desire to make a difference. & \multirow{8}{=}{Desire to
make a difference}\\
Various interests in solving problems and improving lives, often in limited or uncertain fields. \\
Desire to use engineering skills to improve people's lives and make a positive impact. \\
Desire to make a positive impact on society and work for innovative companies. \\
Desire to serve and protect country, often through military or defense-related work. \\
Desire to help others and make a positive impact on the world. \\
Desire to provide for and give back to family, create something, and serve country/community. \\
Interest in creating products that improve people's lives and make a positive impact. \\
Interest in working with prosthetics, improving functionality, and helping people with physical disabilities. \\
Interest in working on impactful, innovative projects and technologies that can change the world. \\
All express a desire to help others through non-profit work or medical aid. \\
Desire to make a positive impact on the world and help people. \\
\botrule

\end{longtable}

Combining this entire process generates something that appears similar to an intermediate step in the codebook generation process during qualitative data analysis. As shown in table \ref{tab-ex-cyb-iss-2-level}, there are top-level codes (second-round summaries), sub-codes (first-round summaries), and examples for the sub-codes (student statements).

\begin{table}[!htb]
\caption{Examples of cybersecurity clusters}\label{tab-ex-cyb-iss-2-level}
\begin{tabularx}{\textwidth}{>{\raggedright\arraybackslash}X >{\raggedright\arraybackslash}X >{\raggedright\arraybackslash}X}
\toprule
Student statements & First-round summaries & Second-round summaries \\
\midrule
Thus, one of my primary goals is to design and program robots that aid
in the extraction of plastic and other types of debris, such as metals,
silicone, and Styrofoam from the ocean because those types of debris are
totally harmful to the marine life. & \multirow{2}{=}{Interest in
environmental conservation, specifically in reducing plastic pollution
and protecting marine life.} & \multirow{5}{=}{Interest in environmental
issues} \\
I really want to work against the use of single-use plastics, so an
example would be reinventing the plastic cup by turning into something
compostable or have it biodegrade quicker than plastic. \\
In high school I enjoyed studying the environment and sustainability. &
\multirow{3}{=}{Interest in addressing environmental issues,
particularly water pollution and sustainability.} \\
I want to decrease the amount of environmental impacts humans have as a
result of our population growth but specific problems that I'm very
passionate in are waste disposal and sustainable energy. \\
\midrule
That's why my ideal position would be a test driver for whatever company
I sign with. & \multirow{3}{=}{Interest in automotive engineering,
designing and testing car parts, improving safety and efficiency.} &
\multirow{6}{=}{Passion for cars and engineering} \\
I would address how to design the car in an appealing way while figuring
out how to also make the car aerodynamically favorable. \\
Some areas of interest include, air-bag development, seat belt
capabilities, and impact research. \\
Working on cars in the past has exposed me to the design elements of all
sorts of cars. & \multirow{3}{=}{Interest in cars and their mechanics,
particularly designing and improving various components.} \\
This field has always been of interest to me, particularly the
mechanical design behind an automobile. \\
\end{tabularx}
\end{table}

\FloatBarrier

\subsection{Deductive Approach: Using Existing Coding Scheme to Analyze
Essays}\label{deductive-approach-using-existing-coding-scheme-to-analyze-essays}

In addition to using the models to identify topics or themes, we also used a priori labels and prompted the model to apply those when analyzing students' responses. In particular, we used a set of labels for identifying career sectors a student may be discussing and a separate set of labels for factors associated with career satisfaction. We then prompted the model to assess its own accuracy of each label it applied to each student statement. The accuracy check results are discussed in the following sections.

\subsubsection{O*NET SOC Labels and Accuracy
Check}\label{onet-soc-labels-and-accuracy-check}

In the first of our two tested deductive approaches, we applied a pre-existing coding scheme based on the O*NET SOC job titles to categorize the students' career sectors mentioned in their essays. A total of 235 job titles were applied to the randomly sampled 100 sentences. This was possible because some statements had more than one label applied to them. For example, a statement such as ``I want to work as a civil or environmental engineer'' would have both ``civil engineer'' and ``environmental engineer'' applied to it. Likewise, a statement such as ``I want to write software for Google'' could have ``software engineer'' and ``software developer'' both applied to it. After analyzing the students' responses and suggesting the applicable label(s), the model assessed its own accuracy. Comparing the model's self-evaluation with the human rater's evaluation of the model's accuracy, we found there were two instances of disagreement (one instance where the model gave a high accuracy rating to an incorrect label and one instance where the model gave a low rating to a label that actually was correct), 219 (93.2\%) instances of agreement (either low accuracy ratings to labels that were indeed incorrectly applied or high accuracy ratings to labels that were correctly applied), and 14 (6\%) instances where the agreement was dubious. The findings of this accuracy check are summarized in table \ref{tab-onet-acc-ck}.

\begin{table}[!htbp]
\caption{Model accuracy check for O*NET SOC job titles}
\label{tab-onet-acc-ck}
\centering
\begin{tabular}{cc}
\toprule
Score category & Count\\
\midrule
Agreement & 219 \\
Disagreement & 2 \\
Questionable & 14 \\
\botrule
\end{tabular}
\end{table}

\begin{figure}[htbp]
\centering
\includegraphics[width=0.8\linewidth]{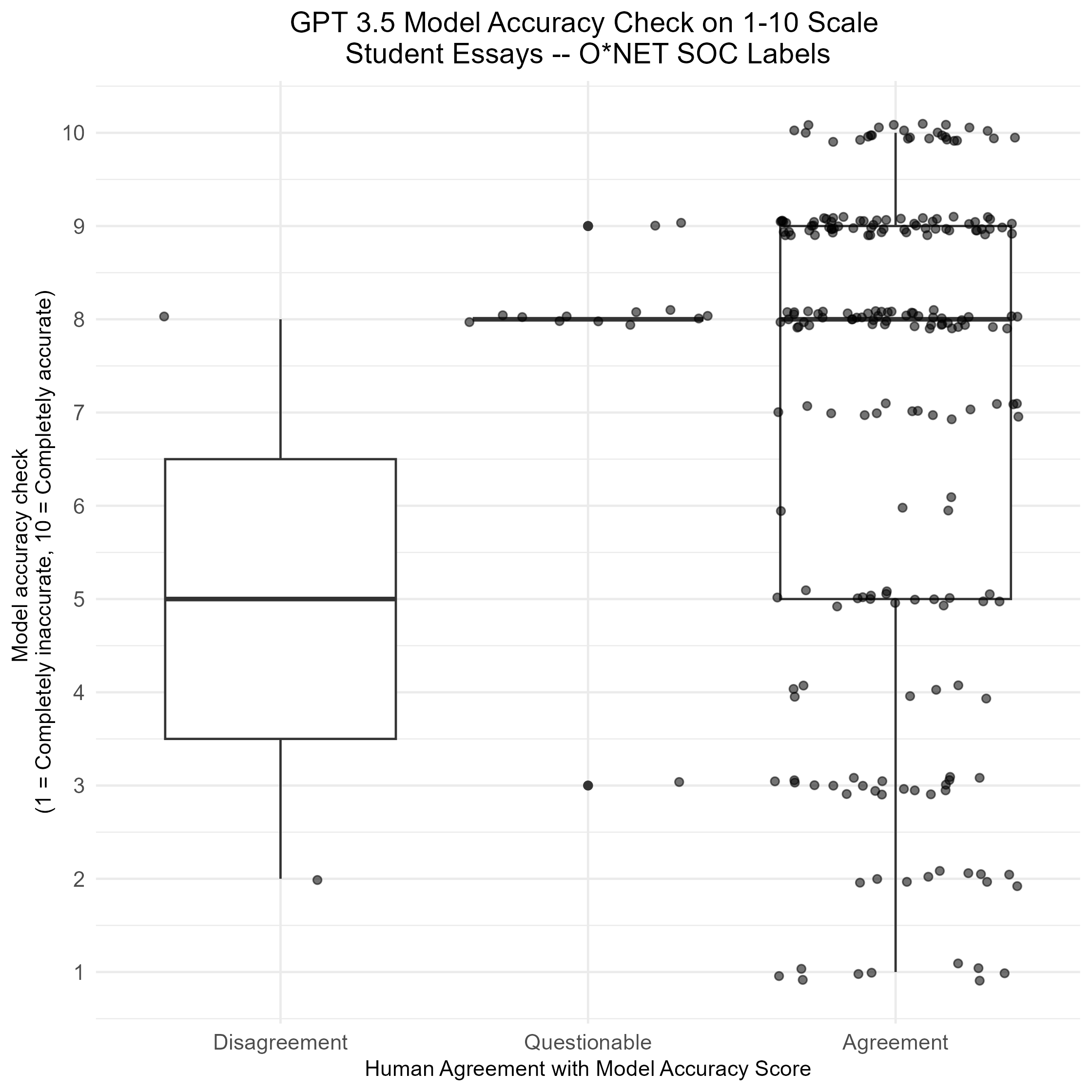}
\caption{Accuracy check figure for O*NET job title labels}\label{fig-onet-ck}
\end{figure}

The numeric accuracy ratings from the model corresponding to each of these three human rating categories are shown in figure \ref{fig-onet-ck}.

These results suggest that some ambiguous or unclear passages in the students\textquotesingle{} essays may have posed challenges for the model in assigning appropriate labels; yet, the model's own ratings of its accuracy correspond with a human rater's assessment of the accuracy
from the initial model's labeling. In this portion of the deductive approach, the model demonstrated a high level of accuracy in applying the O*NET SOC job titles, although the few instances of disagreement and questionable labeling highlight potential areas for improvement and emphasize the importance of human supervision in refining the model's performance. These results also indicate that the first round of labeling by itself may not yield completely accurate results since there were more than 30 instances where the accuracy check model incorrectly suggested a labeled had been misapplied initially (indicated by an accuracy score under 4).

\subsubsection{Career Satisfaction Labeling and Accuracy
Check}\label{career-satisfaction-labeling-and-accuracy-check}

For the second deductive approach, we employed a separate set of labels to identify factors related to career satisfaction in the students' essays. A total of 150 labels were applied to 100 randomly sampled sentences. These instances were assessed in the accuracy check by comparing the model's numeric rating of the label accuracy to the human raters' judgments. The outcomes of this accuracy check are presented in table \ref{tab-car-sat-acc-ck}.

\begin{table}[htbp]
\caption{Model accuracy check for career satisfaction factors}
\label{tab-car-sat-acc-ck}
\centering
\begin{tabular}{cc}
\toprule
Score category & Count\\
\midrule
Agreement & 129 \\
Disagreement & 7 \\
Questionable & 14 \\
\botrule
\end{tabular}
\end{table}

\begin{figure}[htbp]
\centering
\includegraphics[width=0.8\linewidth]{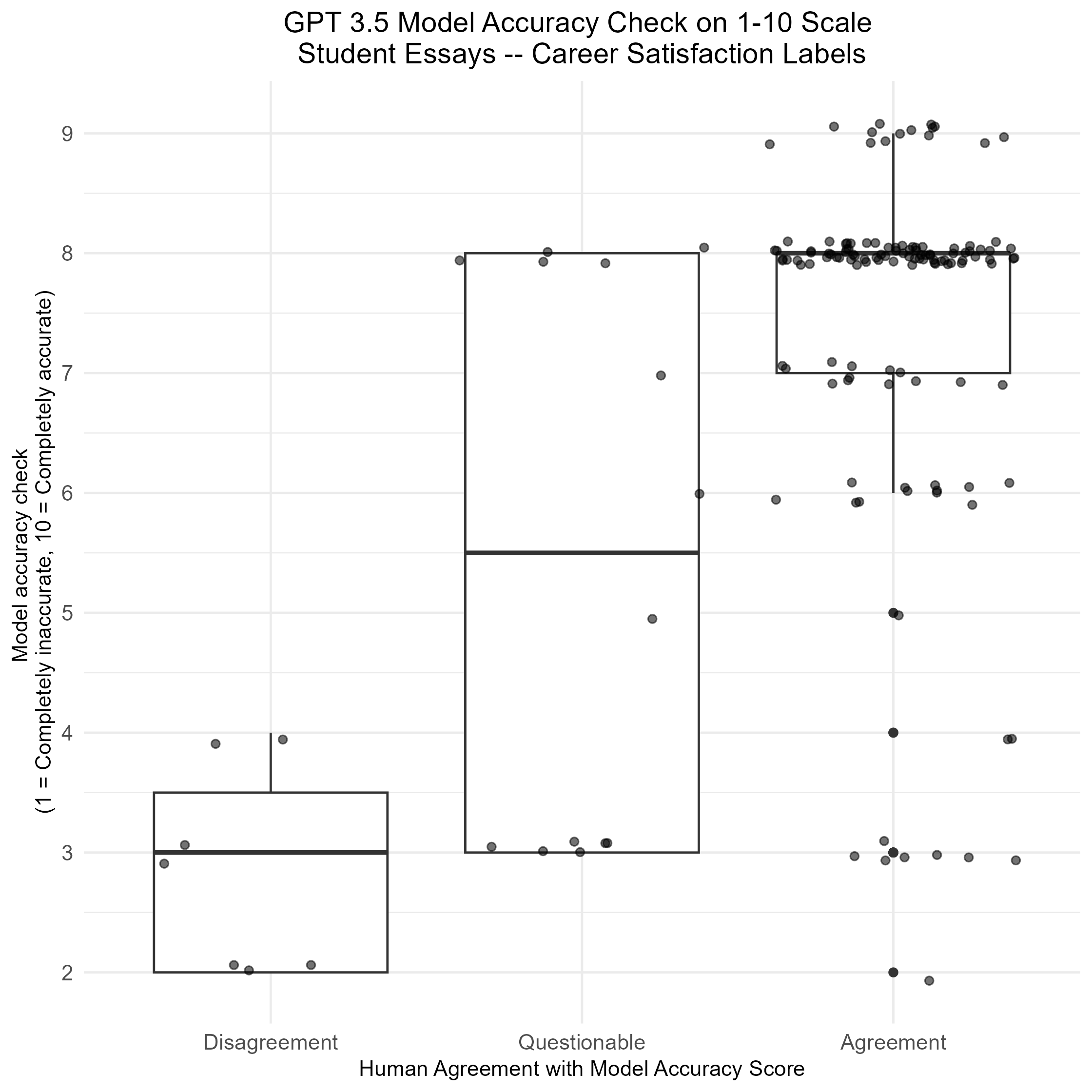}
\caption{Accuracy check figure for career satisfaction labels}\label{fig-car-sat-ck}
\end{figure}

The data in table \ref{tab-car-sat-acc-ck} suggest that the model attained a substantial level of agreement with human raters in 129 instances (86\%). This outcome implies that the model was largely successful in accurately identifying whether or not a label that was suggested during the initial labeling was accurate. Note that instances of agreement simply reflect agreement at the accuracy stage, i.e., if a label was rated at an accuracy of 2 or 7 did the human raters think that was appropriate for judging the accuracy of the initial label. However, there were seven cases (4.67\%) where the model's assigned labels contradicted the human raters' evaluations, indicating areas for potential enhancement in the model's ability to discern specific aspects or contexts within the text. Additionally, there were 14 other instances (9.33\%) where the model's labeling was considered questionable by the human raters. In these cases, the model's label choices were not entirely accurate but not deemed entirely incorrect either, suggesting that some passages in the students' essays presented ambiguity or complexity that challenged the model's ability to assign appropriate labels. The correspondence between the numeric accuracy ratings and the human judgment of the accuracy ratings is shown in figure \ref{fig-car-sat-ck}.

The results of this portion of the deductive approach once again demonstrate a generally high level of accuracy in identifying career satisfaction factors, though the instances of disagreement and questionable labeling continue to underscore the need for human intermittent human supervision. As in the O*NET SOC titles application, the model's self-assessment of its accuracy generally corresponded with the human raters' evaluations, reiterating the significance of combining instruction-tuned LLMs and prompt engineering to label student essays and then screen the accuracy of those labels in a two-step process.

\section{Discussion}\label{discussion}

This study demonstrated several options for analyzing undergraduate engineering student career interest essays. The same variety of techniques can directly be applied in other contexts with other datasets. The results from using these techniques with student career interest essays suggest the combination of text embedding models and text generation models were able to create a preliminary codebook as well as apply pre-defined labels (i.e., O*NET SOC job titles and career satisfaction factors) to student essays with moderately high accuracy but also identified several opportunities for improvement. Specifically, for the inductive approach, there were instances where nuance may be missing with a fully automated approach, such as the difference between wanting to work on sustainability versus working on access to clear water. While these generally fit under the rubric of the UN Sustainable Development Goals \cite{nilsson_policy_2016}, one might want to capture that nuance more distinctly.

With the deductive approach, the accuracy models achieved 93\% and 86\% agreement with human raters in judging the accuracy of the applied O*NET SOC codes and career satisfaction factors, respectively. However, there was also a small percentage of instances (around 5\% in both cases) where the models' labeling was in full disagreement with human judgment. In addition, 6-9\% of labels in each case were considered questionable, where the models' choices were tangentially relevant but not fully precise. At that accuracy level, one would want to have a human in the loop to verify questionable labels. The good news from the accuracy checking step is that the model seems capable of identifying instances of questionable labels, which involves
flagging instances of low accuracy with a low accuracy score (i.e., rating less than 4/10). Therefore, the three-step human-in-the-loop process of (1) model applying a label, (2) model self-evaluating its accuracy, and then (3) a human checking the accuracy of labels with low accuracy scores could improve overall topic labeling accuracy.

\subsection{Broader Lessons of LLMs for Education
Research}\label{broader-lessons-of-llms-for-education-research}

With their continued growth and development, LLMs have demonstrated significant potential for effectively analyzing qualitative data in various domains, including educational research \cite{sok_chatgpt_2023}. The development of these models, such as BERT \cite{devlin_bert_2019} and GPT \cite{radford_language_2019}, as referenced in the Background section, has ushered significant and rapid changes in the field of NLP. By leveraging the power of LLMs, researchers can automate the process of analyzing large volumes of unstructured text data, such as student essays, teacher notes, and interview transcripts \cite{zawacki-richter_systematic_2019}. This capacity arises from the LLMs' ability to capture contextual information embedded within text data \cite{zhang_sentiment_2020}. This capacity allows LLMs to identify patterns and themes that might be difficult to discern through traditional rule-based NLP models and possibly even in manual analysis when datasets become unwieldy \cite{brown_language_2020}. As such, LLMs can significantly reduce the time and effort required for data analysis, enabling researchers to focus on higher-level tasks, such as interpreting the results and formulating actionable insights because of their emergent abilities \cite{wei_emergent_2022}

In addition to the advantages mentioned earlier, LLMs offer a high degree of flexibility in their application to various research questions and data types. For instance, researchers can use LLMs to perform tasks such as clustering \cite{subakti_performance_2022}, summarization \cite{liu_text_2019}, and sentiment analysis \cite{zhang_sentiment_2020} on qualitative data. Furthermore, LLMs can be employed in both inductive and deductive approaches, allowing researchers to adapt the models to their specific research questions and objectives. The present study demonstrated some of these techniques including clustering, summarization, and the combination of inductive and deductive approaches.

The integration of LLMs into educational research not only streamlines the analysis process but also opens up new avenues for exploring student learning and development. For example, by using LLMs to analyze student reflections, educators can gain a deeper understanding of students' thought processes, metacognitive skills, and self-regulation strategies \cite{ghosh_exploratory_2020,wulff_enhancing_2023}. Additionally, LLMs can facilitate the identification of patterns in student behavior, such as engagement, motivation, and collaboration, that may be difficult to detect through traditional quantitative methods \cite{wise_learning_2017}. The important aspect of these developments is the ability to scale analysis in order to identify systematic variations in the data.

As the field of NLP and LLMs continues to evolve, one may expect that these models will become increasingly more powerful and versatile, opening up further possibilities for educational research. For instance, researchers could leverage transformer-based models to analyze multimodal data, such as speech, gestures, and facial expressions, in addition to text \cite{baltrusaitis_multimodal_2018}. The recently released GPT-4 is an example of such a multimodal that can take both text and images as input. Continuing to develop more multimodal models will enable a more comprehensive understanding of students' learning experiences and potentially uncover novel insights into the factors that contribute to their success or failure in educational contexts.

\subsection{Limitations}\label{limitations}

Limitations to the current study come from limitations with the underlying models, including lack of control, bias, and limited capabilities with certain text types. Regarding lack of control, LLMs can generate realistic text but the models remain challenging to control in terms of results, attributes, and constraints \cite{gururangan_dont_2020,tamkin_understanding_2021}. Additionally, bias and toxicity can arise because many of these models are pre-trained on web data, meaning that the LLMs can exhibit bias, toxicity, and other harmful behaviors, inherited from their training data \cite{bender2021dangers}. Fine-tuning and validating on domain-specific data helps address these concerns. Third, LLMs may have difficulty with some text types such as especially from low-resource languages \cite{tapo_neural_2020} due to lack of training data. On that note, with limited data, neural models achieve less progress and encounter ``insufficient learning'' issues \cite{han_pre-trained_2021}. Along with these limitations of the models themselves, there are also non-trivial concerns about the environmental impacts that training and using these models may have \cite{bender2021dangers,rillig_risks_2023,dodge2022measuring}.

Advancing LLMs and explainable AI can improve interpretability and transparency, helping researchers understand and address model limitations. However, LLMs produce complex, data-driven results that can be unexpected or counterintuitive and difficult to control \cite{zellers_defending_2019}. Researchers should validate findings through expert review or other methods to ensure robust, trustworthy conclusions \cite{creswell_qualitative_2016}. This is the motivation in the present study for including a human in the loop. While promising as a tool for gleaning insights from text at a large scale, LLMs still have non-trivial limitations requiring caution and oversight when applying and interpreting their results. Nonetheless, with continued progress we believe that LLMs can gain nuanced insights into student experiences at scale but must still be applied judiciously.

\section{Conclusion}\label{conclusion}

This study demonstrated the promise and limitations of using NLP and LLMs to analyze text data in engineering education research such as undergraduate engineering student career interest essays. The results suggested these techniques can provide nuanced understandings into student experiences at a scale not previously feasible. However, as with any new technology, researchers must exercise caution in their application and interpretation. When combined with human judgment, these methods were able to create a preliminary codebook through embedding and text generation models. The generative models were also able to apply pre-defined labels to student essays with moderately high accuracy. Moreover, the models also showed an ability to identify instances where their labels may be questionable, demonstrating the potential to self-evaluate and provide informative feedback on their own performance. At the same time, there were some cases where nuance seemed to be missing in the models' analyses of student writing, highlighting the need for human oversight, especially for more complex topics. As LLMs become increasingly embedded in digital technology platforms, engineering education researchers have significant opportunities to glean new insights by analyzing large-scale text data sources. One important area for continued research is the use of prompts for interfacing with these generative text models, colloquially called prompt engineering. Documenting best practices for those approaches such as how to phrase a prompt appropriately will further expand the range of what the community can do -- and importantly also recognize the limits of these approaches. Of course, using these generative models also comes with limitations, including difficulty controlling outputs, mitigating biases, and generating certain text types. Fine-tuning models on domain-specific data and validating results can help address these concerns. In addition, advancements in explainable AI may improve the transparency and trustworthiness of LLMs over time. With continued progress, LLMs can become a powerful tool for engineering education research, enabling new discoveries at a scale and depth not previously possible; yet, they must be wielded with caution and care.

\section{References}\label{references}

\bibliography{sn-bibliography}





\backmatter

\section{Acknowledgments}

This material is based upon work supported by the National Science Foundation under Grant No. (2107008).

\begin{appendices}

\section{Prompts for Deductive Approach}\label{app-deductive}

\subsection{Prompts used for labeling with O*NET SOC job titles}\label{app-prompt-onet}
The following is the prompt used for the deductive approach using O*NET SOC job titles.

``The following comment was written in response to the question
\textquotesingle Please describe your future career
interests\textquotesingle. Pick the career from the labels in square
brackets that I provide you that is expressed in the
students\textquotesingle{} written response. If there is more than one
career expressed in the response, separate each with a comma. If no
career from the list is present in the comment, say
\textquotesingle N/A\textquotesingle. CAREERS LIST:
{[}\{labels\_string\}{]}.

STATEMENT:''

Note that labels\_string was a comma delineated string of the 79
engineer job titles above in table \ref{tab-onet-titles}.

For example, the response for the statement ``In addition, when people
bring new code, I may need to test to ensure there are no
vulnerabilities'' was ``Information Security Analysts, Software Quality
Assurance Analysts and Testers''.

\subsection{Prompts used for labeling with career satisfaction factors}\label{app-car-sat-prompt}

The following is the prompt used for the deductive approach using career satisfaction factors.

``The list in square brackets contains factors associated with career
satisfaction.

Career satisfaction factors: {[}\{labels\_string\}{]}.

Does the following comment mention any or these factors? If so, respond
by saying "List: " followed by a list with the factors separated by
commas. If not, write "N/A".

COMMENT:''

Note here that labels\_string was a comma delineated string with the 16
career satisfaction factors listed in table \ref{tab-car-sat-facts}.

For example, the response for the statement ``It also is a company that
pays its engineers very handsomely.'' was ``List: making money''.

\section{Results from all 33 second-level cluster summaries}\label{app-all33}

\begin{longtable}[htb]{p{0.05\linewidth} p{0.55\linewidth} p{0.3\linewidth}}
\caption{All 33 second-level cluster summaries}\label{tab:long} \\
\toprule
 & Clustered statement summaries (first-level summaries) & Second-level summary \\ 
\midrule
\endfirsthead
\multicolumn{3}{c}%
{{\bfseries \tablename\ \thetable{} -- continued from previous page}} \\
\toprule
 & Clustered statement summaries (first-level summaries) & Second-level summary \\ 
\midrule
\endhead
\multicolumn{3}{r}{{Continued on next page}} \\ 
\endfoot
\bottomrule
\endlastfoot
0 & Interest in environmental conservation, specifically in reducing plastic
pollution and protecting marine life.\\
& Interest in careers focused on sustainability, renewable energy, and
environmental protection.\\
& Interest in renewable/sustainable energy and improving energy
efficiency.\\
& Interest in renewable energy and sustainability, including solar and
wind power.\\
& Interest in addressing environmental issues, particularly water
pollution and sustainability.\\
& Interest in using chemical engineering to address environmental and
sustainability issues, including alternative energy sources.\\
& Interest in addressing environmental issues in mining and promoting
sustainability.\\
& Interest in renewable energy and related industries, such as
sustainability and nuclear energy.\\
& Environment and sustainability concerns. & Environmental and sustainability interests. \\
\midrule
1 & 
Interest and passion for coding and computer science as a career
choice.\\
& Interest in pursuing a career in computer science or computer
engineering.\\
& Future career interests involve software development, improving skills,
and working with technology.\\
& Interest in learning and mastering multiple programming languages for
career success.\\
& Interest in math and science, particularly physics and chemistry, and
pursuing a career in STEM.\\
& Interest in pursuing a career in computer science or computer
engineering, often sparked by taking programming classes in high
school.\\
& Interest in computer science, programming, and software development as a
career path.\\
& Future career interests in computer science/engineering, job security,
variety of job opportunities, good pay and benefits.\\
& Future career interests involve software engineering, programming, and
problem-solving.\\
& Future career interests involve computer science, programming, math, and
potential innovation/advancement in the field. & Career interest in technology. \\
\midrule
2 & 
Passion for cars and hands-on experience with cars, often influenced by
family members.\\
& Interest in working as a mechanical engineer in the automotive industry,
specifically for well-known companies or race teams.\\
& Interest in automotive engineering, designing and testing car parts,
improving safety and efficiency.\\
& Interest in working with cars and pursuing a career in the automotive
industry.\\
& Interest in working for car manufacturers, particularly BMW, Mercedes,
Porsche, Ferrari, and Toyota.\\
& Interest in automotive engineering/design, specifically in creating more
efficient and powerful engines and vehicles. & Automotive and roller coaster design. \\
\midrule
3 & Interest in innovation, technology, and problem-solving for future
advancements.\\
& Working in a team to design and develop products or technology, using
CAD and problem-solving skills.\\
& Emphasis on testing, programming, problem-solving, and design.\\
& Interest in problem-solving, logical thinking, and spatial awareness.\\
& Interest in engineering and problem-solving across various industries
and specializations.\\
& Interest in hands-on problem solving, creative thinking, and finding
efficient solutions.\\
& Lack of coherence and relevance to the given prompt.\\
& Interest in hands-on work, design, testing, and innovation in a
project-oriented job.\\
& Emphasis on problem-solving skills in engineering careers.\\
& Interest in programming and technology, concerns about coding
efficiency, complexity, and errors.\\
& Problem-solving is a key skill required in future career
interests. & Problem-solving in innovative careers. \\
\midrule
4 & Interest in healthcare industry, helping others, personal/familial
experiences, desire to make a difference.\\
& Desire to make a positive impact on society and work for innovative
companies.\\
& Desire to serve and protect country, often through military or
defense-related work.\\
& Desire to help others and make a positive impact on the world.\\
& Desire to provide for and give back to family, create something, and
serve country/community.\\
& Interest in creating products that improve people\textquotesingle s
lives and make a positive impact.\\
& Interest in working on impactful, innovative projects and technologies
that can change the world.\\
& All express a desire to help others through non-profit work or medical
aid.\\
& Desire to make a positive impact on the world and help people. & Desire to make a positive impact. \\
\midrule
6 & Interest in using technology to improve medical devices, treatments, and
patient care.\\
& Interest in infectious diseases, prevention, treatment, and research,
particularly in relation to antibiotic resistance.\\
& Interest in biomedical engineering and its impact on healthcare,
technology, and innovation.\\
& Interest in advancing technology for practical use and improving quality
of life.\\
& Interest in designing and developing medical devices to solve clinical
problems and improve quality of life.\\
& Interest in using engineering and technology to improve sports equipment
and athlete performance/safety.\\
& Interest in researching and finding cures for diseases and improving
human life through genetics and biotechnology. & Expert in healthcare technology. \\
\midrule 
7 & Personal experiences and interests driving career aspirations, often
with a desire to make a positive impact.\\
& Desire for personal fulfillment, impact on society, and enjoyment in
chosen career path.\\
& Desiring a job that combines passions, offers variety, and provides
enjoyment and engagement.\\
& Desire for fulfilling, challenging work with personal and global impact,
driven by passion and motivation.\\
& Desire for a fulfilling career that aligns with personal interests and
values.\\
& Future career interests are influenced by high school experiences, work
ethic, and desire for education and job experience.\\
& Interest in sports and business/finance, competitive nature, and diverse
hobbies.\\
& Love for the ocean and water-related activities as a driving force for
career interests.\\
& Positive feelings towards future career interests, including enjoyment,
reward, and excitement.\\
& Searching for a fulfilling career path based on personal interests and
values.\\
& Interest in traveling, experiencing different cultures, and working
abroad. & Career driven by passions. \\
\midrule
8 & Mentioning potential problems or challenges in future career
interests.\\
& Future career interests and goals are mentioned, often related to
specific fields or industries.\\
& Future career interests involve working in teams, facing challenges, and
problem-solving.\\
& Request for a specific career goal that can be planned for.\\
& Describing future career interests and considering various majors and
disciplines.\\
& Future career interests are accompanied by potential challenges and
obstacles.\\
& Interest in addressing various problems in their future career.\\
& Future career interests require academic success, physical fitness, and
passing exams/tests. & Future career interests and challenges. \\
\midrule
9 & All comments discuss the process of discovering and pursuing a specific
career interest.\\
& All comments relate to engineering, design, and innovation in the
automotive industry.\\
& All comments relate to materials, design, and protection in various
fields.\\
& All comments relate to careers in aviation, specifically in piloting or
aircraft maintenance.\\
& All comments relate to career interests in the field of aerospace
engineering.\\
& None of the comments relate to future career interests.\\
& All comments relate to careers in engineering within the military/naval
industry.\\
& All comments relate to careers in ocean or coastal engineering,
including designing structures and managing coastal zones.\\
& All comments relate to careers in civil engineering and construction
management. & Career interests in engineering. \\
\midrule
10 & Future career interests in engineering, requiring specific skills and
education, and often involving materials or electronics.\\
& Engineering offers a broad range of job opportunities in various fields
and industries.\\
& Future career interests in engineering, specifically mechanical,
industrial, and systems engineering, with a focus on working for
established companies.\\
& Career goals involve working in engineering, particularly in the
automotive, aerospace, defense, or technology industries.\\
& Interest in pursuing a career in engineering.\\
& Career goals involve obtaining an engineering degree and often pursuing
further education or specialization.\\
& Interest in engineering careers with high salaries and job growth
projections.\\
& Future career interests in specific companies or industries, often
related to engineering, energy, or law.\\
& Importance of internships, extracurriculars, and networking for career
success in engineering.\\
& Future career interests are focused on engineering, specifically
mechanical, chemical, industrial, and ocean engineering. & Engineering career interests vary. \\
\midrule
11 & Interest in military, defense, and national security-related careers.\\
& Interest in national security, defense industry, and new technologies.\\
& Interest in working for aerospace and defense companies, specifically
Lockheed Martin, for impact on national defense and cutting-edge
technology.\\
& Career goal is to work for a defense/aerospace company, particularly
Lockheed Martin.\\
& Interest in working in the defense industry, specifically in technology
and national security.\\
& Interest in working for government agencies, specifically in military
and intelligence roles.\\
& Interest in working for Boeing.\\
& Interest in careers related to engineering, naval vessels, and defense
industry.\\
& Interest in working for aerospace companies, specifically Lockheed
Martin and Boeing. & Interest in defense careers. \\
\midrule
12 & Future career interests involve learning, growth, and innovation in
various fields.\\
& Uncertainty about future career path, seeking opportunities to explore
and gain experience at Virginia Tech.\\
& Interest in healthcare/medicine, but not necessarily as a doctor or
nurse.\\
& Career interests are focused on engineering, but lack of experience and
interest in specific fields.\\
& Future career interests are related to engineering, internships, and job
opportunities at Virginia Tech.\\
& Uncertainty about specific career interests and openness to various
possibilities.\\
& Uncertainty about specific career goals/majors, but have general ideas
and open to exploration.\\
& Future career interests involve pursuing a career in medicine,
specifically becoming a doctor through medical school. & Career exploration and uncertainty. \\
\midrule
13 & Interest in improving efficiency, reducing waste, and contributing to
the success of products.\\
& Interest in research, laboratory work, and improving products/processes
in various industries.\\
& Interest in cost efficiency, sustainability, and efficiency in design
and production.\\
& Interest in improving manufacturing processes, efficiency, and
developing new products.\\
& Addressing problems in various industries related to quality, safety,
efficiency, and sustainability.\\
& Interest in various industries and their growth potential. & Interest in industrial improvement. \\
\midrule
14 & Future career interests involve overcoming challenges and solving
problems in space exploration and technology.\\
& Interest in working for major aerospace companies such as NASA, SpaceX,
and Boeing.\\
& Interest in working for NASA or a space-related company to contribute to
space exploration.\\
& Interest in working for Tesla, SpaceX, or Elon Musk; focus on renewable
energy, engineering, and innovation.\\
& Interest in space exploration, technology, and potential career
opportunities in the industry.\\
& Interest in space exploration and desire to work for NASA or a
spacecraft design company.\\
& Career interests involve working for NASA or SpaceX, but acknowledge
challenges and competition.\\
& Interest in working on spacecraft, satellites, rovers, and other
mechanical devices for space exploration.\\
& Career interest in working for space-related companies such as NASA and
SpaceX. & Space career aspirations. \\
\midrule
15 & Interest in pursuing a career in chemical engineering or related fields,
with a focus on innovation and problem-solving.\\
& Interest in engineering, proficiency in CAD software, technical and
mathematical skills, communication abilities.\\
& Interest and experience in drafting, CAD, and engineering design. & Engineering and design interests. \\
\midrule
16 & Interest in management positions, specifically in engineering or project
management.\\
& Interest in working for civil/environmental engineering firms or service organizations like Engineers Without Borders.\\
& Interest in engineering, specifically ocean and civil engineering, with
some consideration for environmental engineering.\\
& Interest in engineering management, overseeing projects, and
environmental impact.\\
& Interest in civil engineering, specifically in designing and
constructing structures such as buildings, bridges, and roads.\\
& Interest in pursuing a career in civil engineering, specifically in
designing and building infrastructure projects.\\
& Future career interests related to environmental engineering and
law. & Engineering and management interests. \\
\midrule
17 & Interest in game development, creating immersive experiences, addressing
glitches, and improving game mechanics.\\
& Future career interests involve technology and gaming industries.\\
& Interest in pursuing a career in the video game industry as a developer
or designer.\\
& Interest in virtual reality technology and improving its realism and
accessibility.\\
& Interest in video games as a medium for storytelling, community
building, and personal growth.\\
& Interest and passion for video games since childhood, desire to create
or contribute to game development.\\
& Interest in game design and development, including programming, art, and
mechanics. & Passionate about game development. \\
\midrule
18 & All express interest in leadership and management roles.\\
& Interest in managerial/supervisory roles, efficiency, project
management, coding, and teamwork.\\
& Interest in working collaboratively in a team environment towards a
common goal.\\
& Focus on managing projects, teams, and resources to achieve efficiency
and productivity.\\
& Interest in leadership and desire to be in charge of a team or
company.\\
& Emphasis on technical skills, communication, teamwork, problem-solving,
and ability to work under pressure.\\
& Overseeing projects, managing budgets, ensuring deadlines are met, and
distributing tasks. & Leadership and management interests. \\
\midrule
19 & Descriptions of various career opportunities and ways to gain experience
in different industries.\\
& Future career interests involve gaining work experience through
internships/co-ops in desired fields.\\
& Interest in internships, job shadowing, and exploring various career
paths and industries.\\
& Desiring to enhance skills and experience through extracurriculars,
internships, and academic achievements.\\
& Focus on gaining experience, skills, and moving up the career ladder.\\
& Need for work experience, education, communication skills, and
competitive edge to obtain desired job.\\
& Work experience is a key requirement for future career interests. & Career experience is crucial. \\
\midrule
20 & Interest in naval architecture, shipbuilding, and solving related
problems.\\
& Interest in naval/marine engineering, designing ships, and addressing
mechanical/system failures.\\
& Interest in naval engineering and technology, influenced by family
members and experiences. & Naval engineering interests highlighted. \\
\midrule
22 & Interest in working with robotics, AI, and autonomous machines to solve
problems and improve efficiency.\\
& Interest in artificial intelligence and machine learning, specifically
in software development and research.\\
& Interest and experience in robotics and programming, involvement in
robotics teams and competitions.\\
& Interest in artificial intelligence and machine learning for future
career.\\
& Interest in robotics, automation, and artificial intelligence in various
fields and applications.\\
& Interest in robotics and/or related fields, desire to create and
innovate, past experience in robotics.\\
& Interest in artificial intelligence and machine learning, desire to work
on cutting-edge technology and make breakthroughs. & Expert in robotics and AI. \\
\midrule
23 & Interest and fascination in a particular field or subject.\\
& Interest in specific classes/topics and enjoyment in learning them.\\
& Focus on learning through reading and experience, networking, and
communication skills. & Passionate learning through experience. \\
\midrule
24 & Interest in technology, computers, electronics, and building/fixing
things.\\
& Interest in improving technology, solving problems, and benefiting
others through computer engineering.\\
& Interest in technology, specifically computer hardware and software
development.\\
& Interest in computer hardware, technology, and engineering.\\
& Interest in understanding and creating complex machinery, particularly
in the field of engineering and computer science.\\
& Interest in software development/engineering, programming, and creating
innovative technology/products. & Interest in computer engineering. \\
\midrule
25 & Interest in aircraft engineering, design, and innovation.\\
& Interest in designing and improving space travel technology and
systems.\\
& Interest in working with drones, particularly for military and delivery
purposes.\\
& Interest in improving propulsion technologies for more efficient and
sustainable space travel.\\
& Interest in aviation, planes, and military aircraft.\\
& Interest in pursuing a career in aerospace or related engineering
fields, with a focus on innovation and impact.\\
& Interest in aerospace engineering, rocket design, propulsion systems,
and space travel.\\
& Interest in aerodynamics and flight, specifically aircraft and
spacecraft design and engineering.\\
& Interest in aerospace engineering, improving aerodynamics, and
efficiency of aircraft and spacecraft.\\
& Interest in addressing problems related to space exploration and
aerospace engineering. & Aerospace engineering and innovation. \\
\midrule
26 & Interest in owning and running own business/startup, often in
engineering/tech field.\\
& Interest in working for reputable, established, and well-known companies
with resources and opportunities for growth.\\
& Interest in good employee benefits, high job satisfaction, and
innovative companies with modern work environments.\\
& Difficulty in getting a job at highly competitive and well-known
companies.\\
& Concerns about company size, work intensity, and personal freedom.\\
& Comparison of job security and benefits between private sector and
government jobs.\\
& Desiring to start own business or have control over career direction and
impact.\\
& Desiring to work for a specific company due to its reputation, location,
and/or industry. & Career interests and concerns. \\
\midrule
27 & Interest and fascination with space and space exploration since
childhood.\\
& Interest in space exploration and discovery, potential for improving
life on Earth.\\
& Interest in space exploration, curiosity about the universe, desire to
expand knowledge and develop technology.\\
& Interest in space exploration, specifically working towards getting
humans to Mars and advancing technology.\\
& Interest in Mars colonization as a solution to Earth\textquotesingle s
environmental problems and human survival. & Passion for space exploration. \\
\midrule
28 & 
Career goals require determination, hard work, and serving others to
make a positive impact.\\
& Describing a personal goal and the reasons for wanting to achieve it.\\
& Describing a long-term goal that requires hard work, dedication, and
overcoming obstacles. & Goals require hard work. \\
\midrule
29 & 
Interest in working for large, well-known technology companies such as
Microsoft, Google, Apple, and Intel.\\
& Interest in working for Google as a software developer/engineer due to
its innovation, diversity, and reputation.\\
& Interest in working for Google due to its reputation, resources,
innovation, and opportunities.\\
& Interest in AI, machine learning, data science, and working for tech
giants/companies like Amazon, Google, and Facebook.\\
& Desire to work for Google as a software engineer or in a technological
role.\\
& Interest in working for large technology companies such as Apple and
Microsoft. & Tech giants job interest. \\
\midrule
30 & 
Family influence and inspiration in career interests.\\
& Career interests are driven by financial stability and high salary.\\
& Family members or acquaintances in engineering influenced career
interests.\\
& Family influence on career interests.\\
& Career interests are driven by company values, impact on society, and
personal fulfillment.\\
& Career aspirations influenced by family members, particularly fathers,
in business and tech fields. & Family shapes career interests. \\
\midrule
31 & Addressing problems related to efficiency, technology, and software
development.\\
& Interest in using technology to solve complex problems and work with
large amounts of data.\\
& Problem-solving skills and use of data to improve performance and
prevent issues in various industries.\\
& Interest in creating software/apps to make life easier and more
efficient on a large scale. & Expert in tech problem-solving. \\
\midrule
32 & Interest in engineering, building, and creating projects involving
technology and materials.\\
& Interest in designing and creating products using technical skills such
as engineering, physics, and programming. \\
& Interest in creative, technical, and innovative careers in film,
engineering, and design.\\
& Interest in technology, engineering, problem-solving, and creativity.\\
& Interest in building and creating, often starting with Legos or
childhood tinkering.\\
& Interest in creativity, innovation, and inventing new products or
ideas. & Interest in technical creativity. \\
\bottomrule
\end{longtable}




\end{appendices}


\end{document}